\documentclass[twocolumn,secnumarabic,amssymb, nobibnotes, aps, prl, superscriptaddress]{revtex4-1}

\usepackage{natbib}
\usepackage{graphicx}
\usepackage{amsmath}
\usepackage{empheq}
\newcommand{\EXP}[1]{\textrm{e}^{\displaystyle #1}}

\usepackage{tabularx}

\usepackage{graphicx}
\usepackage{color}
\usepackage{enumerate}
\usepackage[latin1]{inputenc} 

\usepackage{lineno}

\begin{document}

\title{Nonlinear chirped Doppler interferometry for $\chi^{(3)}$ spectroscopy}

\author{Elizaveta Neradovskaia}
\affiliation{Institut de Physique de Nice (INPHYNI), Universit\'e C\^{o}te d'Azur, CNRS, UMR 7010, 1361 route des Lucioles, 06560 Valbonne, France}

\author{Benjamin Maingot}
\affiliation{Institut de Physique de Nice (INPHYNI), Universit\'e C\^{o}te d'Azur, CNRS, UMR 7010, 1361 route des Lucioles, 06560 Valbonne, France}
\affiliation{Fastlite, 165 route des Cistes, 06600 Antibes, France}

\author{Gilles Ch\'eriaux}
\affiliation{Institut de Physique de Nice (INPHYNI), Universit\'e C\^{o}te d'Azur, CNRS, UMR 7010, 1361 route des Lucioles, 06560 Valbonne, France}

\author{Cyrille Claudet}
\affiliation{Institut de Physique de Nice (INPHYNI), Universit\'e C\^{o}te d'Azur, CNRS, UMR 7010, 1361 route des Lucioles, 06560 Valbonne, France}

\author{Nicolas Forget}
\affiliation{Fastlite, 165 route des Cistes, 06600 Antibes, France}

\author{Aur\'elie Jullien}
\email{Aurelie.Jullien@inphyni.cnrs.fr}
\affiliation{Institut de Physique de Nice (INPHYNI), Universit\'e C\^{o}te d'Azur, CNRS, UMR 7010, 1361 route des Lucioles, 06560 Valbonne, France}

\begin{abstract}

Four-wave mixing processes are ubiquitous in ultrafast optics and the determination of the coefficients of the $\chi^{(3)}$ tensor is thus essential. We introduce a novel time-resolved ultrafast spectroscopic method to characterize the third-order nonlinearity on the femtosecond time-scale.
This approach, coined as "nonlinear chirped Doppler interferometry", makes use of the variation of the optical group delay of a transmitted probe under the effect of an intense pump pulse in the nonlinear medium of interest. The observable is the spectral interference between the probe and a reference pulse sampled upstream. We show that the detected signal is enhanced when the pulses are chirped, and that, although interferometric, the method is immune to environmental phase fluctuations and drifts. By chirping adequately the reference pulse, the transient frequency shift of the probe pulses is also detected in the time domain and the detected nonlinear signal is enhanced. Nonlinear phase shifts as low as  10\,mrad, corresponding to a frequency shift of 30\,GHz, i.e. 0.01\% of the carrier frequency, are detected without heterodyne detection or active phase-stabilization. The diagonal and/or non-diagonal terms of reference glasses (SiO$_2$) and crystals (Al$_2$O$_3$, BaF$_2$, CaF$_2$) are characterized. The method is finally applied to measure the soft vibration mode of KTiOAsO$_4$ (KTA).
\end{abstract}

\maketitle

\section{Introduction}

The third-order susceptibility tensor $\chi^{(3)}$ governs both resonant and non-resonant nonlinear processes involving four-wave mixing \cite{Sutherland:2003aa}. The real part of the diagonal terms, responsible for the well-known optical Kerr-effect (OKE), plays a fundamental role in ultrafast optics where frequency-degenerated nonlinear processes are routinely used to modify, control or characterize the spatial, spectral and temporal properties of intense pulses. The non-diagonal terms of the tensor $\chi^{(3)}$ are also exploited in nonlinear processes such as cross-polarized wave generation \cite{Jullien200510-10-temporal-} and are ubiquitous when intense waves of different polarizations are mixed in nonlinear media. This is, in particular, the case of most optical parametric amplifiers, which have become the backbone of third-generation femtosecond sources \cite{Fattahi:2014aa}. It is thus essential to determine the coefficients of the $\chi^{(3)}$ tensor of nonlinear media ranging from common optical glasses to exotic nonlinear birefringent crystals. 

Experimental methods suitable to measure the real part of the $\chi^{(3)}$ tensor are numerous \cite{Sutherland:2003aa} and suffice it to say that OKE spectroscopy gathers time-resolved pump-probe techniques tracking the changes (polarization, frequency, spatial phase or temporal phase) of a weak probe pulses under the effect of an intense pump pulse in the medium of interest, either without \cite{DeSalvo1996Infrared-to-ult,Chien:00,Chen:07,Borzsonyi:10, Burgin2005Femtosecond-inv} or with \cite{LeBlanc:00,Thurston:2020aa} interferometric detection. Another related approach, two-beam coupling (2BC), relies on the mutual interaction between two noncollinear beams crossing in the medium \cite{Kang:97,Smolorz:00,Smolorz:1999vf, Wahlstrand:2013wl, Patwardhan:21}. All these methods have their own advantages and drawbacks, but most of them require noise reduction to isolate the contribution of weak nonlinearities (averaging and/or modulation with heterodyne detection).

In this paper, we report a novel time-resolved ultrafast transient spectroscopy method to characterize  third-order nonlinearity on the femtosecond time-scale. We coin this method as "nonlinear chirped Doppler interferometry". As explained in the next section, "Doppler" refers to the observable, "interferometry" describes the detection method and "chirped" points out the fact that adequately chirped pulses are required.

The approach consists in monitoring the variation of the optical group delay of a transmitted probe under the effect of an intense pump pulse, instead of the phase changes. 
We demonstrate that, under certain chirp conditions, two distinct physical effects (spectral and temporal shifts) add up and that monitoring the optical group delay, via spectral interferometry, gives access to the nonlinear phase value and therefore to the nonlinear tensor terms of the medium of interest. We show, both theoretically and experimentally, that the method is:
\begin{enumerate}
    \item interferometric but immune to environmental phase fluctuations and drifts: no active stabilization or shielding of the interferometer is required,
    \item highly sensitive: nonlinear phase-shifts as low as 10\,mrad can be detected without heterodyne detection,
    \item selective: self-focusing although visible on the experimental data does not affect the measurement,
    \item polarization sensitive: non-diagonal terms of the $\chi^{(3)}$ tensor can be independently measured,
    \item temporally-resolved: both instantaneous and delayed nonlinear processes can be investigated.
\end{enumerate}
The paper is organized as follows. The theoretical line-out of nonlinear chirped Doppler interferometry is described in section 2. The experimental setup and nonlinear materials are then described in section 3. Section 4 presents measurement results : validation of the technique with known isotropic and $\chi^{(3)}$-anisotropic materials, followed by soft vibration mode measurement in an anisotropic crystal (KTA).


\section{Principle}

The setup is essentially a frequency-degenerated pump-probe experiment with an interferometric detection. As justified below, the probe optical group delay as a function of pump-probe delay is monitored, rather than the phase. 

\begin{figure*}[htp]
\centering
\includegraphics[width=2\columnwidth]{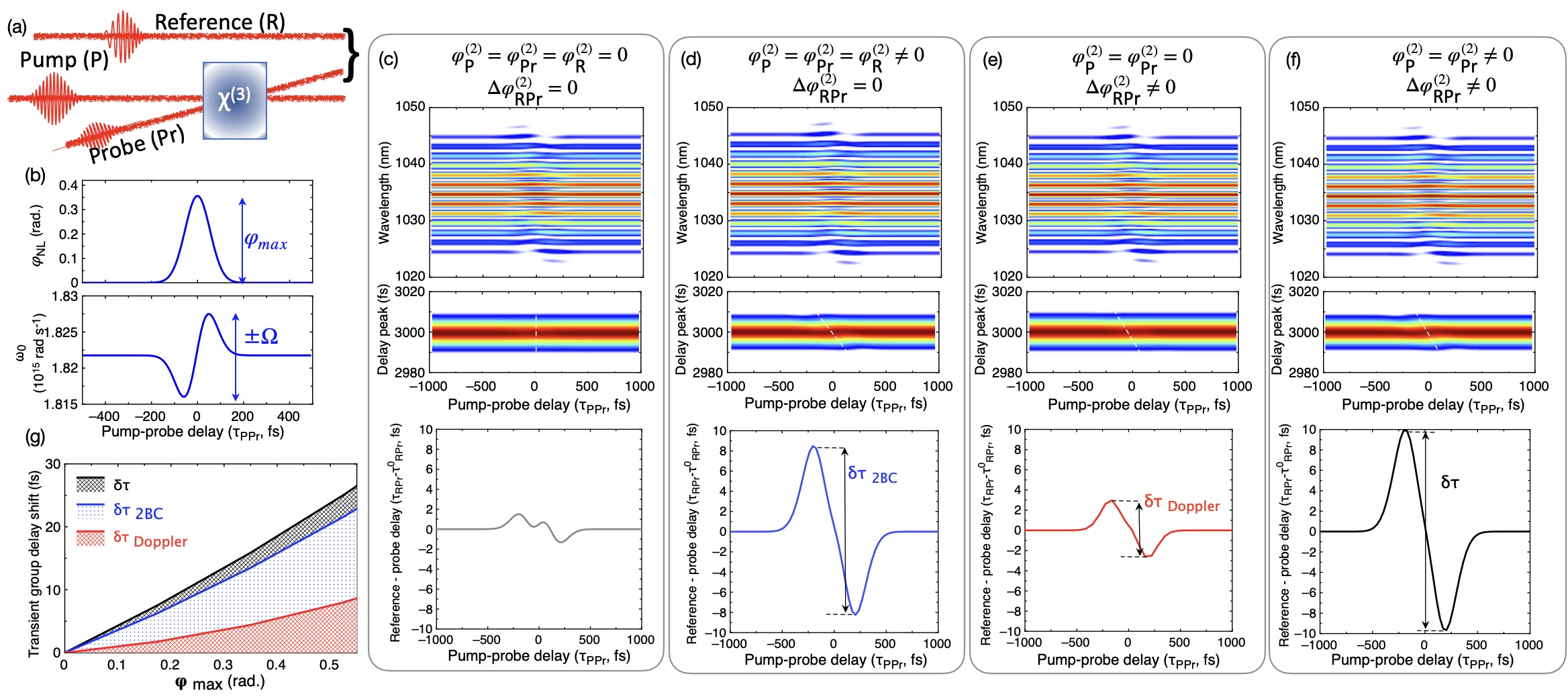}
\caption{
(a) Principle of nonlinear Doppler chirped interferometry. 
(b-g) Numerical illustration for the numerical values indicated in the text.
(b) Nonlinear XPM phase and shift of the carrier frequency of the transmitted probe pulse as a function of the pump-probe delay $\tau_\text{PPr}$.
Top row of (c,d,e,f): computed spectral interference $S(\omega)$ between the reference and the transmitted probe as a function of $\tau_\text{PPr}$. Middle row of (c,d,e,f) : modulus of the Fourier transform $|\hat{S}_\text{AC}(t-\tau_\text{RPr}^0)|$ as a function of $\tau_\text{PPr}$. Bottom row of (c,d,e,f): group delay shift of the transmitted probe $\tau_\text{RPr}-\tau_\text{RPr}^0$ as a function of $\tau_\text{PPr}$. The four columns correspond to the following cases: all three pulses are unchirped (c), all pulses are chirped by the same amount (d), only reference pulse is chirped (e), all pulses are chirped with an additional chirp on the reference pulse (f). 
(g) Amplitude of the transient shift ($\delta\tau$) of the probe group delay as a function of the XPM nonlinear phase for the three cases (d - blue line; e - red line; f - black line).
}
\label{fig:principe}
\end{figure*}

An illustration of the principle is available in Fig. \ref{fig:principe}a. A strong pump beam ($P$) and a weak probe beam ($Pr$), of identical carrier frequency $\omega_0$, are weakly focused and cross in a thin optical sample. A small angle ($\simeq 1^{\circ}$) is introduced between the two beams to allow a spatial separation before/after the sample and a delay line controls the relative group delay between the two pulses ($\tau_\text{PPr}$). Through cross-phase modulation (XPM), the probe pulse undergoes a transient nonlinear phase shift $\varphi_\text{NL}(t)$. The temporal dependence of $\varphi_\text{NL}$ causes a shift of the probe's instantaneous carrier frequency, $\pm\Omega$ (Fig. \ref{fig:principe}b): an down-shift (shift toward the "red") on the rising edge of the pump pulse and a up-shift (shift toward the "blue") on the trailing edge of the pump pulse (for a medium characterized by a positive nonlinear index $n_2>0$) \cite{Boyd1992Nonlinear-optic}. $\Omega$ is proportional to the nonlinear phase, and thus to the involved $\chi^{(3)}$ term \cite{Boyd1992Nonlinear-optic}. An interpretation of this effect relies on the fact that, because of the nonlinear index of refraction, the probe pulse propagates in a medium of increasing/decreasing optical thickness. This is as if the radiation source were moving away from or toward the observer, causing an (unrelativistic) Doppler effect on the observed probe frequency. For this reason, the frequency shift $\Omega$ will be referred to as a the nonlinear Doppler shift hereafter.
In nonlinear spectroscopy, $\varphi_{NL}$ can be extracted from the transient variation of $\Omega$, although with little precision \cite{Ripoche:1997aa}.
We assume in the following that this frequency shift, noted $\Omega=\Omega(\tau_{PPr})$, is small compared to the optical bandwidth $\Delta\omega$ such that  $|\Omega|<<\Delta\omega$.

Let $E_\text{P}(\omega)$, $E_\text{R}(\omega)$ and $E_\text{Pr}(\omega)$ be the complex spectral amplitudes of, respectively, the pump, reference and probe pulses. Their respective  spectral phases are $\varphi_\text{P}(\omega)$, $\varphi_\text{R}(\omega)$, $\varphi_\text{Pr}(\omega)$ with the convention $E_k(\omega) = |E_k(\omega)|\exp\left[-i\varphi_k(\omega)\right]$ where $k$ is the wave label. Spectral phases are hereafter assumed as purely quadratic, with chirp coefficients respectively labeled  $\varphi^{(2)}_{k}$. The spectral phases thus write 
\begin{equation}\label{eq:taylor}
    \varphi_k(\omega) = \varphi_k(\omega_0) + \tau_k(\omega_0) (\omega-\omega_0) + \varphi^{(2)}_{k} (\omega-\omega_0)^2/2
\end{equation}
where $\tau_k = \tau_k(\omega_0)$ stands for the group delay. 

The coupling between the pump and probe beams is treated elsewhere, and can be described by the following propagation equation for the probe field \cite{Wahlstrand:2013wl} in the limit of a purely electronic nonlinearity: 
\begin{equation}\label{eq:xpm}
    c\frac{\partial E_\text{Pr}}{\partial z}+\left(n_{g,0}+4\gamma I_\text{P}\right)\frac{\partial E_\text{Pr}}{\partial t} = 2 i\omega_0 \gamma I_\text{P}E_\text{Pr}-4\gamma\frac{\partial I_\text{P}}{\partial t}E_\text{Pr}
\end{equation}
where $I_\text{P}$ is the time-dependent pump intensity, $n_{g,0}$ the group index at $\omega_0$, and $\gamma$ is the nonlinear coupling coefficient. This expression holds as long as the input polarization states of the pump and probe beams are either parallel or perpendicular with respect to each other. The general expression of $\gamma$ is rather complex and depends on the polarization states of the pump and probe pulses as well as on the crystallographic orientation and symmetry of the sample. For an isotropic medium, far away from any resonance, the expressions of $\gamma$ in SI units, for respectively parallel and perpendicular polarizations, are:
\begin{align}\label{eq:n2}
    &\gamma_\parallel = \frac{3}{4\epsilon_0 n_0^2 c}\chi^{(3)}_{xxxx} = n_2\\
    &\gamma_\perp = \frac{1}{4\epsilon_0 n_0^2 c}\chi^{(3)}_{yyyy}
\end{align}
The propagation equation Eq~\ref{eq:xpm} assumes slowly varying envelopes and neglects temporal dispersion which is compatible with the thin medium assumption and/or narrowband pulses. The effect of the pump field is three-fold: the group index is increased by $4\gamma I_\text{P}$ (left member of Eq.\ref{eq:xpm}), and two nonlinear source terms contribute to the propagation (right member of Eq.\ref{eq:xpm}). The first source term corresponds to XPM and is responsible for the Doppler frequency shift $\Omega$, while the second term induces gain and loss via an energy transfer between the two beams (2BC). The interplay between these three effects is rather complex but, to simplify, the change in wave velocity can be neglected while the XPM and 2BC may significantly reshape the transmitted probe pulse in both the spectral and time domains. As a general result, the optical group delay of transmitted probe pulse $\tau_\text{Pr}$ is altered when the pump and probe pulses overlap. For the sake of clarity, the probe group delays with and without the pump beam are respectively noted $\tau_\text{Pr}$ and $\tau_\text{Pr}^0$.

As $\tau_\text{Pr}-\tau_\text{Pr}^0$ cannot be measured directly, the transmitted probe pulse is recombined with a reference pulse (in this case a replica selected upstream) and the relative group delay $\tau_\text{RPr}=\tau_\text{R}-\tau_\text{Pr}$ between reference and probe is measured instead. We note $\tau_\text{RPr}^0 = \tau_\text{R}-\tau_\text{Pr}^0$ so that $\tau_\text{Pr}-\tau_\text{Pr}^0 = \tau_\text{RPr} - \tau_\text{RPr}^0$. The relative delay $\tau_\text{RPr}^0$, which is kept constant during the experiment, is chosen adequately so as to be able to resolve the phase difference between the probe and reference pulses by spectral interferometry: $\Delta\omega \ll 1/|\tau^0_\text{RPr}| < \delta\omega_\text{sp}$ where $\delta\omega_\text{sp}$ is the spectral resolution of the spectrometer. For plane waves, the spectral interference pattern between the transmitted probe and the reference writes:
\begin{multline}\label{eq:signal}
    S(\omega) = |E_\text{R}(\omega)|^2 + |E_\text{Pr}(\omega)|^2 \\ + 2\text{Re}\left\{ E_\text{R}(\omega) E_\text{Pr}^*(\omega)\EXP{i\omega\tau_{RPr}}\right\}
\end{multline}
$S(\omega)$ contains one non-oscillating term (DC term) and two conjugate oscillating terms (AC terms), the Fourier-transform of which is:
\begin{equation}\label{eq:FT}
    \hat{S}_\text{AC}(t-\tau_0) = \int E_\text{R}(\omega)E_\text{Pr}^*(\omega)\EXP{i\omega t}d\omega
\end{equation}
If $E_\text{Pr}(\omega) = E_\text{R}(\omega)$ (ie case, without the pump wave), then $\hat{S}_\text{AC}(t-\tau_\text{RPr}^0)$ is equal to the Fourier-transform of $|E_\text{R}(\omega)|^2$, and the AC terms are centered at $t=\pm\tau_\text{RPr}^0$ and well separated from the DC term at $t=0$. To anticipate on  the following paragraphs, four-wave mixing introduces additional contributions to the optical group delay of the probe pulse and tends to shift the location of the AC terms from $\pm\tau_\text{RPr}^0$ to $\pm\tau_\text{RPr}$ (our observable). As shown below, this definition actually aggregates two distinct physical effects.\\

A numerical resolution of Eq.\ref{eq:xpm}, described in the S.M., is proposed for a pump pulse duration of 180\,fs FWHM, $I_\text{P}$ = 300\,GW/cm$^2$ and a crystal of 1\,mm length characterized by a nonlinear index $\gamma=n_2\text{ = 2.8 10}^{-16} \text{ cm}^2\text{/W}$. The corresponding nonlinear XPM phase is $\simeq$300\,mrad. 
Pump and probe pulses are equally chirped, with a chirp coefficient of either $\varphi^{(2)}_\text{P}$ = $\varphi^{(2)}_\text{Pr}$ = 0\,fs$^2$, or $\varphi^{(2)}_\text{P} \text{ = } \varphi^{(2)}_\text{Pr}\text{ = +5000 fs }^2$. Reference pulse has a chirp coefficient of $\varphi^{(2)}_\text{R}$ = $\varphi^{(2)}_\text{Pr}$ + $\Delta\varphi^{(2)}_\text{RPr}$, with either $\Delta\varphi^{(2)}_\text{RPr}$ = 0\,fs$^2$ or $\Delta\varphi^{(2)}_\text{RPr}$ = +2000\,fs$^2$. 
The initial group delay between the probe and the reference pulses is $\tau_\text{Pr}^0$=3\,ps. As our toy model is unidirectional, spatial effects such as Kerr lens and self-diffraction are not simulated.

We first consider the case $\varphi^{(2)}_\text{P}=\varphi^{(2)}_\text{R}=\varphi^{(2)}_\text{Pr}\text{ = 0 fs }^2$, i.e. all involved pulses are limited by Fourier transform. 
The spectral interferogram as a function of pump-probe delay, $\tau_\text{PPr}=\tau_\text{P}-\tau^0_\text{Pr}$, is plotted in Fig.~\ref{fig:principe}c. The Doppler transient spectral shift\,$\Omega$ appears for $\Delta\Omega\tau_{PPr} \lesssim 1$, when the pump and probe pulses temporally overlap. 
For each pump-probe delay, the discrete Fourier transform of the spectral interferogram is computed and the relative group delay between the reference and probe pulses ($\tau_{RPr}$) is retrieved by fitting the AC peak $|\hat{S}_\text{AC}(t-\tau_\text{RPr}^0)|$ with a Gaussian function. As shown on Fig.~\ref{fig:principe}c, $\tau_\text{RPr}-\tau_\text{RPr}^0$ varies with the pump-probe delay $\tau_\text{PPr}$. When none of the three pulses are chirped, weak variations are observed, indicating that, to the first order, the optical group delay of the transmitted probe pulse is constant despite the spectral/temporal reshaping effects.

\begin{figure*}[ht!]
\centering
\includegraphics[width=2\columnwidth]{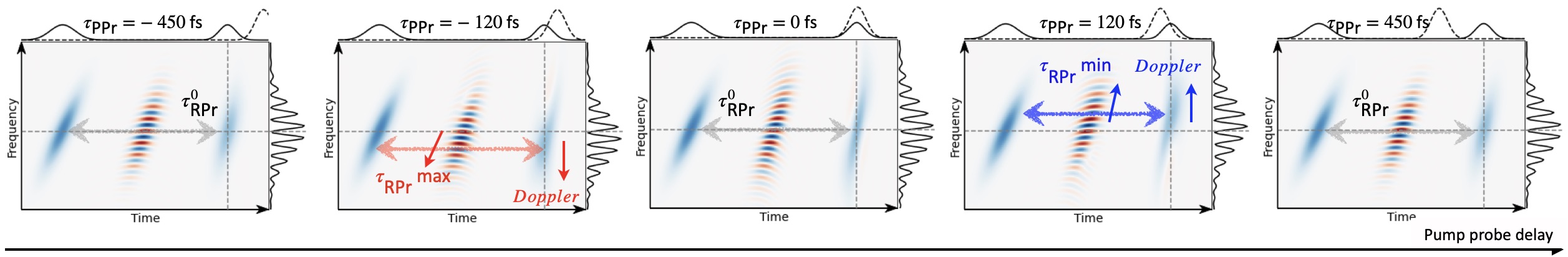}
\caption{Wigner-Ville distribution of the probe-reference electric fields for five different pump-probe delays $\tau_\text{PPr}$. 
For each sub-plot, the reference pulse is on the left, and the (delayed) probe pulse is on the right. Both pulses are chirped and exhibit a time-dependent instantaneous frequency. The two pulses being chirped differently, the slopes of the individual representations differ. In between the two pulses appears the interferometric component. Upper plot: integration of the distribution along the frequency coordinate (temporal intensity, solid line) compared to the cross-section of the pump pulse (dotted line). Right plot: integration of the distribution along the time coordinate (interference spectrum $S(\omega)$). 
The large arrow indicates the reference-probe group delay $\tau_\text{RPr}$.
For $\tau_\text{PPr}=\pm \text{450 fs}$ (i.e. no temporal overlap between pump and probe) $\tau_\text{RPr}=\tau_\text{RPr}^0$. For $\tau_\text{PPr}\lesssim 0$, the nonlinear phase induces a red shift of the transmitted probe (red arrow), which increases $\tau_\text{RPr}$. Symmetrically, for $\tau_\text{PPr}\gtrsim 0$, a blue shift decreases $\tau_\text{RPr}$. For $\tau_\text{PPr}=0$, there is no spectral shift and $\tau_\text{RPr}=\tau_\text{RPr}^0$. The nonlinear phase and chirp values were increased compared to Fig.\ref{fig:principe} for the sake of the illustration.}
\label{fig:wigner}
\end{figure*}

We then consider $\varphi^{(2)}_\text{P}=\varphi^{(2)}_\text{R}=\varphi^{(2)}_\text{Pr}\text{ = +5000 fs }^2$, i.e. the three pulses are equally chirped. Because of this chirp, the instantaneous frequencies of the pump and probe pulses are detuned with respect to each other when $\tau_\text{PPr}\neq 0$ and energy flows from one wave to the other during the nonlinear interaction (2BC becomes dominant). If the chirp coefficient is positive, the probe will gain energy for negative pump-probe delays and vice versa. As this energy transfer also scales with the pump intensity, the general effect is a reshaping of the temporal profile which is indistinguishable from an additional optical group delay (Fig.~\ref{fig:principe}d). For negative pump-probe delays (resp. positive), the rear edge (resp. the leading edge) of the probe is strengthened, resulting in a overall increase (resp. decrease) of the group delay. As a result $\tau_\text{RPr}$ exhibits a Z-shape, similar the transient probe transmission reported for two-beam coupling \cite{Kang:97,Smolorz:00,Smolorz:1999vf, Patwardhan:21}. 

We now consider the case of unchirped pump and probe pulses $\varphi^{(2)}_\text{P}=\varphi^{(2)}_\text{Pr}\text{ = 0 fs }^2$ with a positively chirped reference pulse $\varphi^{(2)}_\text{R}\text{ = +2000 fs }^2$ (Fig.~\ref{fig:principe}e). The reference pulse being chirped, the Doppler spectral shift $\Omega$ is temporally encoded in the interferogram and (also) appears as a delay $\tau_\text{RPr}$. As plotted in Fig.~\ref{fig:principe}e, this effect produces a similar Z-shape behaviour, although less pronounced than in the former case - but scaling linearly with $\Delta \varphi^{(2)}_\text{RPr}=
\varphi^{(2)}_\text{R}-\varphi^{(2)}_\text{Pr}$. To illustrate the principle of spectral encoding, we represent in Fig.~\ref{fig:wigner} the Wigner-Ville distributions of the transmitted probe and of the delayed and chirped reference pulse. As developed in the supplementary material, the linear relationship between the nonlinear Doppler shift ($\Omega$) and the relative chirp between probe and reference ($\Delta \varphi^{(2)}_\text{RPr}$) can be retrieved analytically from Eq.\ref{eq:FT}.  

We have evidenced here two phenomena: temporal reshaping and temporal encoding of the nonlinear spectral shift. They originate from different mechanisms but are the two sides of the same coin. Although the differences between the spectrograms in Fig. \ref{fig:principe} are not visible to the naked eye, the Fourier analysis shows that both mechanisms result in similar transient shifts of $\tau_\text{RPr}$, over the same temporal scale (the correlation width of the pump pulse), and with similar amplitudes (a few fs). With the right chirp parameters, these two contributions may add up and increase significantly the global signal-to-noise ratio of the measurement, as shown in Fig.~\ref{fig:principe}f. The delay swing $\delta\tau=\max(\tau_\text{RPr})-\min(\tau_\text{RPr})$ is then evolving linearly with the nonlinear phase amount, as plotted in Fig.~\ref{fig:principe}g. As will be demonstrated experimentally in the next section, measuring $\tau_\text{RPr}$ instead of phase changes not only makes the detection insensitive to phase fluctuations but also gives additional means to enhance the sensitivity and specificity (Fig.~\ref{fig:principe}g) of the detection, without resorting to heterodyne detection.


\section{Experimental methods}

\begin{figure}[h]
\centering
\includegraphics[width=1\columnwidth]{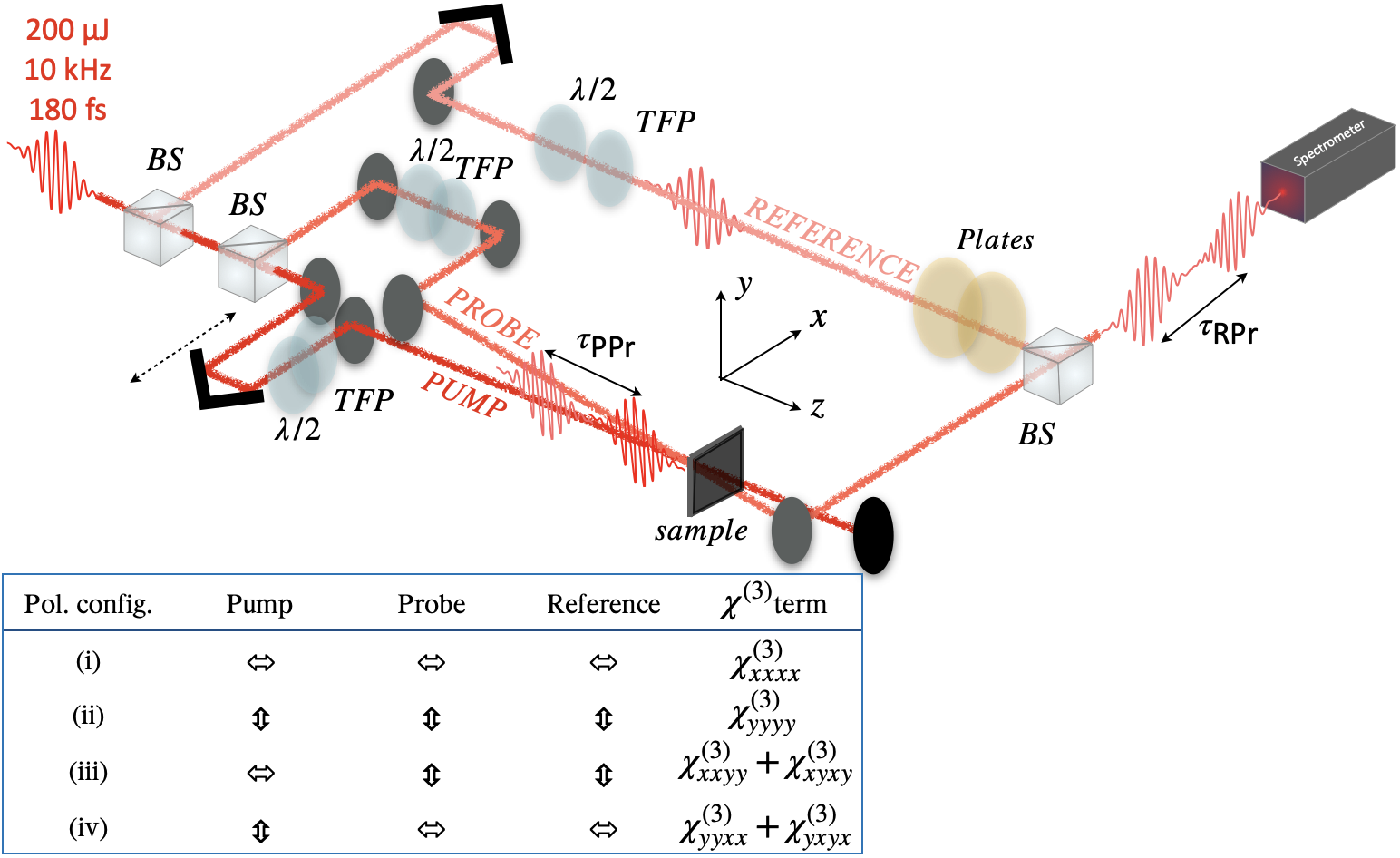}
\caption{
Optical setup. The laser source (Pharos SP, Light Conversion) is split into three beams (pump, probe, and reference).
BS: beam splitter. TFP: thin film polarizer. $\lambda/2$: half-wave plates.
Pump and probe beams are focused ($f$=1.5\,m, not shown) on the nonlinear sample, positioned on a 3-axis translation stage.
Optional glass plates are inserted in the reference beam path.
The table summarizes the polarization configurations and measured components of $\chi^{(3)}$ tensor.}
\label{fig:setup}
\end{figure}

In the present experiments, a Pharos laser system (PH1-SP-1mJ, Light Conversion) delivers 180\,fs FTL pulses, with a central wavelength of 1034\,nm, a repetition rate of 10\,kHz, and pulse energy up to 500\, $\mu$J. The laser chirp can be tuned by adjusting the compressor. Each pulse is split into three separated pulses: an excitation pump pulse, a probe pulse ($\simeq$20\% of the pump energy), and a reference pulse selected before the nonlinear stage (Fig.~\ref{fig:setup}).
The pump and probe pulses are focused ($\text{f = 1.5 m}$) and overlap in the focal plane under a small angle  with a pump beam size of about 550\, $\mu$m. The pump-probe delay ($\tau_\text{PPr}$) is controlled with a delay stage equipped by a motorized actuator with K-Cube controller (models Z825B and KDC101, Thorlabs). 
After the interaction, the probe is selected and recombined with the reference pulse. The nominal group delay between the reference and the probe pulses ($\tau_\text{RPr}^0$) is set to 3.5$\pm$0.15\,ps for all measurements. The two pulses, probe and reference, are respectively chirped through the addition of various bulk plates: SF11 (20 mm, $\text{2500 fs}^2$), CaCO$_\text{3}$(10 mm, $\text{430 fs}^2$)\cite{Ghosh:99} and Al$_\text{2}$O$_\text{3}$ (5 mm, $\text{156 fs}^2$ )\cite{Malitson:72}. The resulting interference pattern is collected by a spectrometer (Avantes, spectral resolution 0.07\,nm) for each step of the optical delay stage in the pump arm.
The polarization and energy of each pulse are controlled by half-wave plates and thin-films polarizers (TFP). The different components of the $\chi^{(3)}$ tensor can then be measured by changing the polarization state of the three pulses. 

The detailed procedure for data acquisition and analysis can be found in the S.M. To summarize, for each acquisition scan, $\tau_\text{PPr}$ is scanned (single scan) from -2.6\,ps to 2.6\,ps with temporal steps of 13\,fs (400 spectra per scan). The integration time of the spectrometer is 1\,ms (total acquisition time <1\,mn, limited by the delay stage). The probe-reference delay and relative chirp are measured before each acquisition, out of pump-probe temporal overlap. $\tau_\text{RPr}$ is computed by fitting the modulus of the AC peak with a Gaussian function after discrete Fourier transform. The relative phase between the probe and reference pulses was also retrieved by Fourier filtering, so as to compare our analysis with usual nonlinear phase measurements. 

The method was validated against a set of isotropic materials for which the nonlinear refractive indices are well known : fused silica (1 mm, $\text{(2.2}\pm \text{0.3)}\times\text{10}^{-16} \text{ cm}^2\text{/W}$ at 1030 nm) \cite{Kabacinski:19}, barium fluoride (0.5 mm, $\text{(2.2}\pm \text{0.2)}\times\text{10}^{-16} \text{ cm}^2\text{/W}$ at 1064 nm) \cite{Adair1989nonlinear-refra}, sapphire (1 mm, $\text{(2.8}\pm \text{0.7)}\times\text{10}^{-16} \text{ cm}^2\text{/W}$ at 1030 nm) \cite{Major:04} and calcium fluoride (1 mm, $\text{(1.2}\pm \text{0.3)}\times\text{10}^{-16} \text{ cm}^2\text{/W}$ at 1030 nm) \cite{Kabacinski:19}. All plates are uncoated. In order to check the validity of various polarization configurations, we also characterized a barium fluoride crystal, with holographic $[011]$ crystallographic orientation, which is known to present an anisotropy of its third-order nonlinearity, $\sigma$, defined as follows \cite{Minkovski2004Nonlinear-polar,Canova2008Efficient-gener}: 
\begin{equation}
\sigma = \frac{\chi^3_{xxxx}-2\chi^3_{xyyx}-\chi^3_{xxyy}}{\chi^3_{xxxx}} 
\label{sigma}
\end{equation}

\section{Results}

The method is validated in three steps. We first compare the toy model with experimental data and measure the swing amplitude of $\tau_\text{RPr}-\tau_\text{RPr}^0$ with respect to the relative chirps and nonlinear phase. We then measure the diagonal third-order tensor coefficient of a set of well-known isotropic samples. Last, we vary the polarization states of the pump and probe beams, characterize the nonlinear anisotropy of barium fluoride and compare the measured coefficients with the reported values. Finally, we investigate the instantaneous and delayed nonlinear properties of potassium titanyl arsenate crystal (KTA).

\subsection{Comparison with the toy model}

\begin{figure}[htp]
\centering
\includegraphics[width=1\columnwidth]{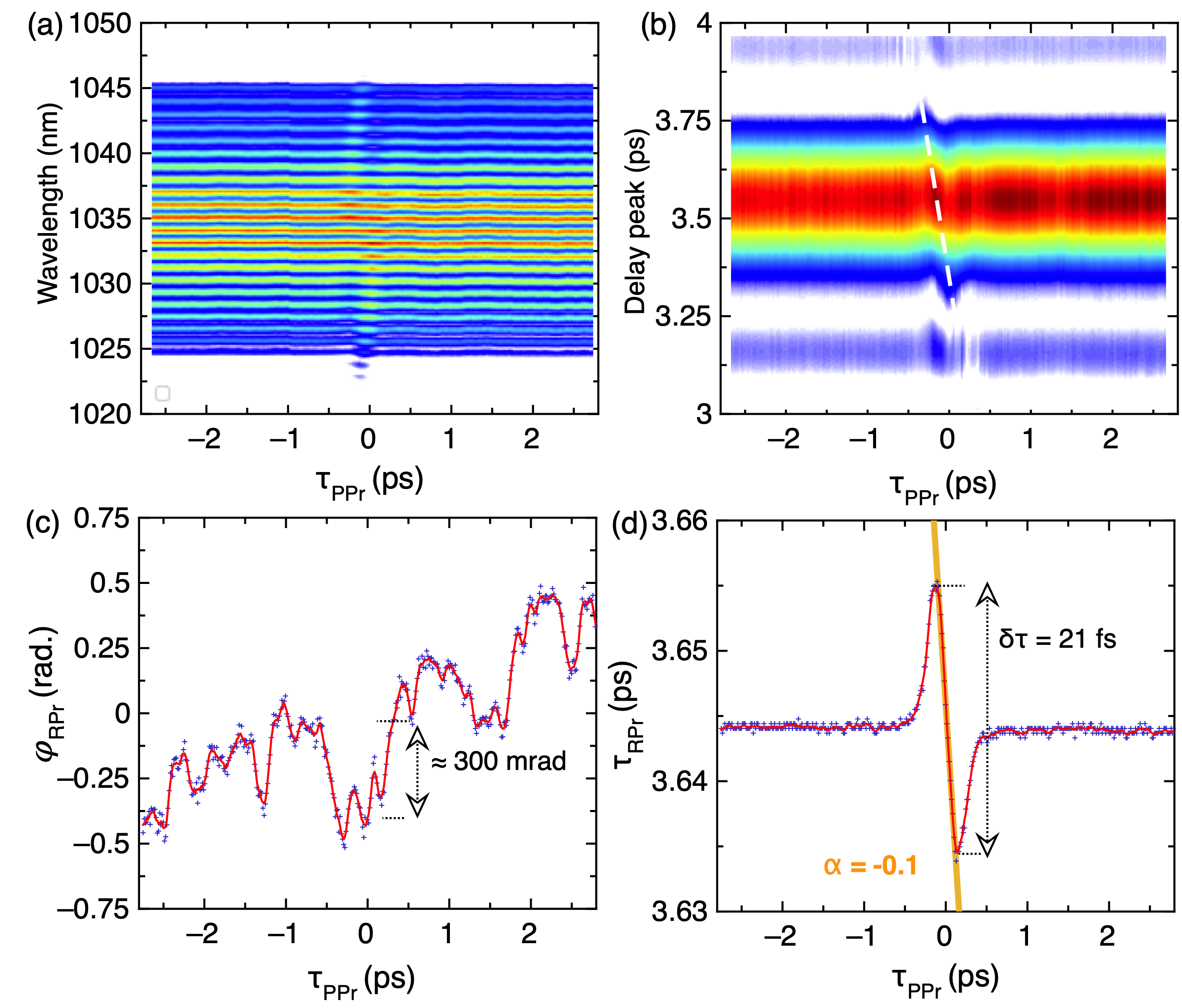}
\caption{Typical experimental data for the 1-mm-thick sapphire. The nonlinear phase-shift accumulated by the probe is estimated to 300 mrad and the Doppler shift is $\Omega/\omega_0\simeq10^{-3}$.  
(a) Interferogram of the probe and reference pulses as a function of pump-probe delay.
(b) AC peak of the Fourier transform of the interferograms. The white dashed line underlines the linear temporal shift of the delay.
(c) Relative phase between reference and probe pulses ($\varphi_\text{RPr}$) as a function of pump-probe delay. 
(d) $\tau_\text{RPr}$ as a function of pump-probe delay.
For sub-plots (c-d) : raw data (blue), moving average (red), linear fit (orange).
}
\label{fig:data}
\end{figure}

All polarization directions are first set to horizontal. The pump energy is 40\, $\mu$J, which corresponds to a pump peak intensity estimated to 300\,GW/cm$^2$ on the sample. The measured chirp coefficients are $\varphi^{(2)}_\text{P} = \varphi^{(2)}_\text{Pr} \text{ = (+5000}\pm\text{500) fs}^2$ and $\Delta\varphi^{(2)}_\text{RPr}\text{ = (+2000}\pm\text{100) fs}^2$. Fig.~\ref{fig:data} gathers typical experimental results acquired for a 1\,mm c-cut sapphire crystal. As expected (Fig. \ref{fig:data}a), in the vicinity of $\tau_\text{PPr}=0$, a red-shift is observed for $\tau_\text{PPr}<0$ while a blue-shift is observed for $\tau_\text{PPr}>0$. Self-focusing is visible on the experimental data as a transient signal decrease but does not affect the measurement. 
A close-up of the Fourier transform near +3.5\,ps is shown in Fig. \ref{fig:data}b. The transient shift of the peak position ($\tau_\text{RPr}$) is clearly distinguishable. The latter, plotted in Fig. \ref{fig:data}d, shows the characteristic z-shape, in excellent agreement with Fig.~\ref{fig:principe}. The delay swing $\delta\tau=\max(\tau_\text{RPr})-\min(\tau_\text{RPr})$\,=\,21\,fs is obtained for $\tau_\text{PPr}=\pm\text{100 fs}$ with a linear dependence of the signal between these two extrema (slope of $\simeq$-0.1). For the sake of comparison, Fig.~\ref{fig:data}c represents the relative phase between the reference and the transmitted probe pulses extracted from the spectrogram by Fourier filtering. This metric is the quantity usually measured to determine the value of the non-linear phase. However, as can be seen on Fig.~\ref{fig:data}c, $\varphi_\text{NL}$ can hardly be distinguished from the phase noise (fluctuations and drifts) added by the interferometer. The comparison between figures \ref{fig:data}c,d helps to appreciate the improvement in signal-to-noise ratio of our method. Monitoring the group delay rather than the spectral phase is indeed insensitive to phase fluctuations and makes it possible to measure a nonlinear signal at least an order of magnitude times weaker.

\begin{figure}[htp]
\centering
\includegraphics[width=1\columnwidth]{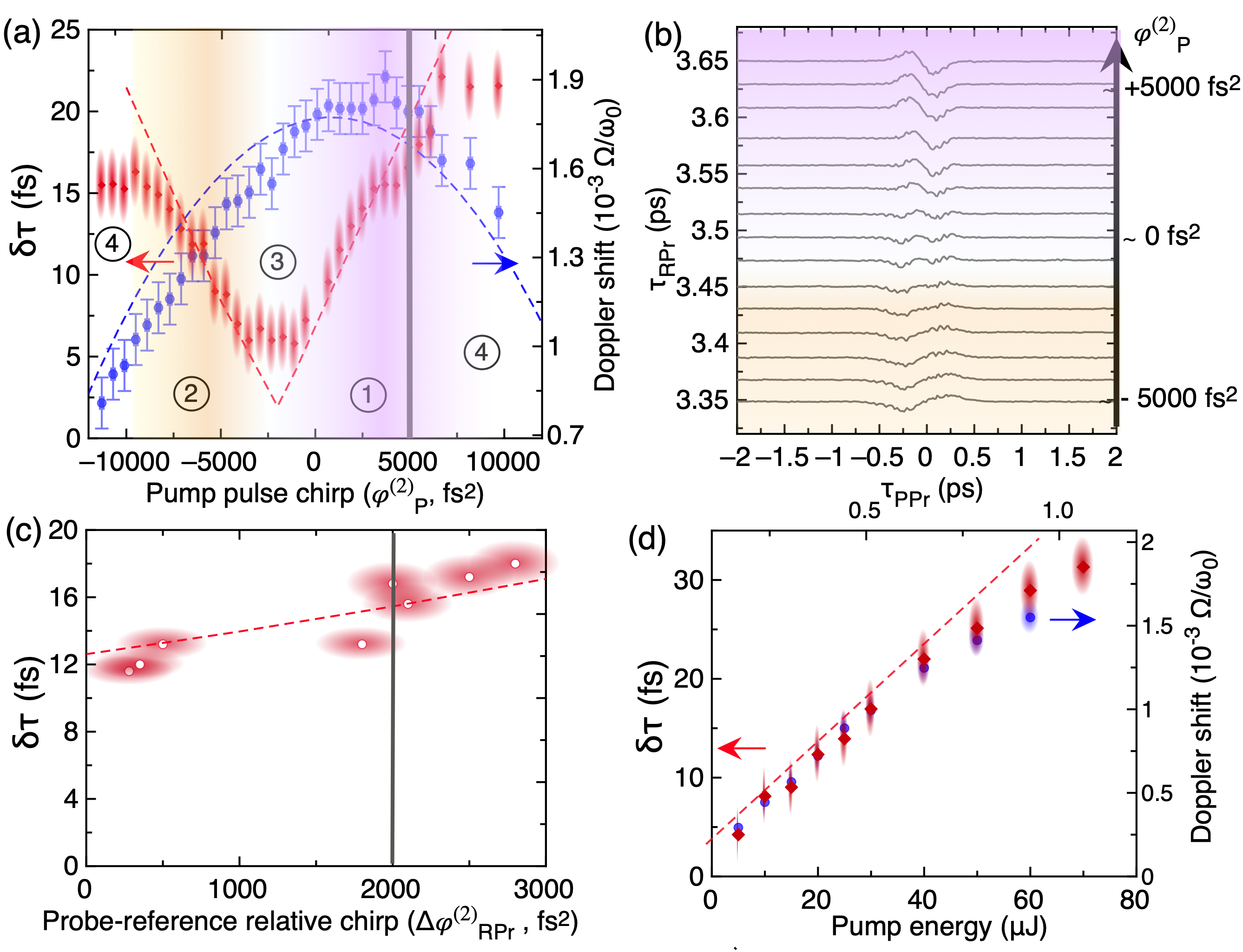}
\caption{
Experimental data compared to the numerical 1D simulations. The clouds indicate the dispersion of the experimental data.
(a) $\delta\tau$ (red) and normalized Doppler shift ($\Omega/\omega_0$, blue) as a function of input chirp $\varphi^{(2)}_\text{P}=\varphi^{(2)}_\text{Pr}$, for $\Delta\varphi^{(2)}_\text{RPr}\text{ = (+2000}\pm\text{100) fs}^2$. Colored area and labels refer to different zones of interest (see text). 
(b) $\tau_\text{RPr}$ as a function of $\tau_\text{PPr}$, for various $\varphi^{(2)}_\text{P}$ values, indicated on the right y-scale. The curves are artificially shifted for clarity. The colored area correspond to those in (a). 
(c) $\tau_\text{RPr}$ as a function of the chirp difference between the reference and pump/probe pulses, $\Delta\varphi^{(2)}_\text{RPr}$, with fixed input chirp $\varphi^{(2)}_\text{P} \text{= (+5000}\pm\text{500) fs}^2$.
(g) $\tau_\text{RPr}$ as a function of the pump energy for the chirp values indicated by the gray lines in plots (a) and (c).
}
\label{fig:chirp}
\end{figure}

Starting from this working configuration, we then characterize $\delta\tau$ as a function of (i) input chirp (controlled by the grating compressor), (ii) $\Delta\varphi^{(2)}_\text{RPr}$ (controlled by adding/removing bulk in the reference beam path) and, (iii) pump energy - all other experimental parameters being kept constant. The acquired raw data are shown in Fig.~\ref{fig:chirp}. As explained in section 2, the input chirp mainly triggers 2BC, meanwhile $\Delta\varphi^{(2)}_\text{RPr}$ tunes temporal encoding of the Doppler shift.
The toy model reproduces well the variations of $\tau_\text{RPr}$ and $\Omega$ (normalized with respect to $\omega_0$) with the input chirp (Fig. \ref{fig:chirp}a).  The Doppler shift follows the parabolic evolution of pump intensity, with a maximum close to pulse compression. Conversely, $\delta\tau$ varies mostly linearly (V-shape) around a minimum, staggered from the compression. 
We can distinguish several trends, as labelled in Fig. \ref{fig:chirp}a.
For moderate positive input chirp (1), the Doppler shift remains mostly constant (i.e. so is the nonlinear phase) while $\delta\tau$ exhibits a linear increase with the chirp value, emphasizing how the signal to noise ratio is enhanced by 2BC. This area then defines the efficient working conditions of our method. For negative input chirp (2), similar trend is observed, but with a reversed Z-shape, as shown in the raw data in Fig. \ref{fig:chirp}b. Reversing the sign of the chirp is in fact equivalent to modify the sign of the energy couplings between the pump and the probe and thus to reverse the signal. 
The minimum of $\delta\tau$ V-shape (3) corresponds to a weak signal with ill-defined shape (Fig. \ref{fig:chirp}b). The large offset from pulse compression is both predicted by the model and observed. It matches the reference-probe relative chirp, -2000\,fs$^2$, and results from the compensation of the temporal and spectral contributions detailed before. 
The numerical fit agrees quite well with the experimental data, except for the large chirps (4). This discrepancy is attributed to higher orders of spectral phase (neglected in the model).

The increase of $\delta\tau$ with $\Delta\varphi^{(2)}_\text{RPr}$ is shown in Fig.~\ref{fig:chirp}c: the detected signal varies linearly with the relative chirp value, even if the pump energy remains the same. This additional degree of freedom, easy to implement experimentally, is be particularly useful to increase $\delta\tau$ in order to detect a weak amount of nonlinear phase. 
Finally, we characterize how $\delta\tau$ and $\Omega$ scale with the pump intensity by scanning the pump energy from 5\, $\mu$J and 80\, $\mu$J. Fig. \ref{fig:chirp}d evidences the linear dependence of both $\delta\tau$ and $\Omega$ with the pump energy up 50\, $\mu$J. This measurement also exemplifies that Doppler shifts spanning over more than an order of magnitude (from 30\,GHz to 400\,GHz, i.e. from 0.01\% to 0.2\% of the carrier frequency) can be detected in a single scan.

To summarize this sub-section: (i) experimental data are found in very good agreement with the toy model, (ii) the expected trends of the signal dependence on involved chirp parameters have been recovered, (iii) suitable experimental chirps and pump energy ranges so as to monitor nonlinear phase changes have been identified. 

\subsection{Application to isotropic and anisotropic crystals}

Although the method can in principle provide the absolute value of the nonlinear index, experimental sources of error are numerous (actual peak intensity, uncertainties on chirps...) and, in this work, we characterize $\delta \tau$ relatively to a reference sample, hereafter fused silica. The experimental parameters are those shown in Fig. \ref{fig:data}. 
Table \ref{tab:table2_v2} gathers the measured $\chi^{(3)}$ ratio between the samples to characterize (Al$_\text{2}$O$_\text{3}$, BaF$_\text{2}$, CaF$_\text{2}$) and fused silica ($\chi^{(3)}_0$). A good agreement with published data ($\chi^{(3)}_{lit}$) is found. It can be noted that the error interval is similar or lower to the ones given in the literature. 

The last step of this validation process consists in changing the set of polarizations to check the symmetry properties of the nonlinear tensor. 
Comparison between polarization configuration $(i)$, $(ii)$ and $(iii)$ enables to verify the tensor symmetry for fused silica : $\chi^{(3)}_{0,xxxx}=\chi^{(3)}_{0,yyyy}$ and $\chi^{(3)}_{0,xxyy}=\chi^{(3)}_{0,yyxx}\simeq{\chi^{(3)}_{0,xxxx}}/{3}$. 
Then, BaF$_\text{2}$ with holographic orientation was characterized. The ratio between  $\delta \tau$ obtained for two distinct polarization configuration leads to a determination of the ratio between SPM and XPM $\chi^{(3)}$ terms, and thus to the nonlinear anisotropy of $\chi^{(3)}$ ($\sigma$, eq. \ref{sigma}) \cite{Minkovski2004Nonlinear-polar, Canova2008Efficient-gener,Kourtev:09}. Our results are summarized in Table \ref{tab:baf2}, and once again, a good agreement between the values found in the literature is obtained. 

\begin{table}
\begin{tabular}{lcc}
\hline
Sample & $\chi^{(3)}/\chi_{0}$ (this work)& $\chi^{(3)}_{lit}/\chi_{0}$ (literature)\\
\hline
Al$_\text{2}$O$_\text{3}$& $(1.1\pm 0.2)$ & $(1.3\pm 0.3)$ \cite{Major:04} \\
BaF$_\text{2}$& $(1.0 \pm 0.1)$ & $(1.0\pm 0.1)$ \cite{Adair1989nonlinear-refra} \\
CaF$_\text{2}$ &$(0.5 \pm 0.02)$ &$(0.6\pm 0.1)$ \cite{Kabacinski:19}\\
\hline
\end{tabular}
\caption{\label{tab:table2_v2}  $\chi^{(3)}$ ratio between validation set plates, normalized to fused silica nonlinear term hereafter labeled as $\chi_{0}$. $\chi_{0}=(2.0\pm0.2)\,10^{-22}\,\text{m}^{2}/\text{V}^2$ \cite{Gubler:00}. }
\end{table}

\begin{table}
\begin{tabular}{cccc}
\hline
Polarization configuration & $\sigma$ (this work) & $\sigma$ \cite{Minkovski2004Nonlinear-polar}\\
\hline
(i),(ii) & $-1.32\pm 0.1$& $-1.15\pm 0.1$\\
(i),(iii) & $-1.42\pm 0.1$& $-1.15\pm 0.1$\\
\hline
\end{tabular}
\caption{\label{tab:baf2} Nonlinear anisotropy of BaF$_\text{2}$ with [011] crystallographic orientation.}
\end{table}

Finally, we study the nonlinear properties of a Potassium Titanyl Arsenate (KTA) crystal along $\theta$=47$^{\circ}$ and $\phi$=0$^{\circ}$  (X-Z plane, thickness of 2\,mm). The pump energy is reduced to 15 $\mu$J, so as to keep the nonlinear phase within the linearity range indicated in Fig. \ref{fig:chirp}d. The polarization configurations are successively set to (i) and (ii), to measure the nonlinear indices of the fast and slow axes. Our results are gathered in Tab. \ref{tab:table_kta} and found slightly lower than previous measurements reported using the Z-scan technique \cite{Li:2001tm}. 

\begin{figure}[h]
\centering
\includegraphics[width=1\columnwidth]{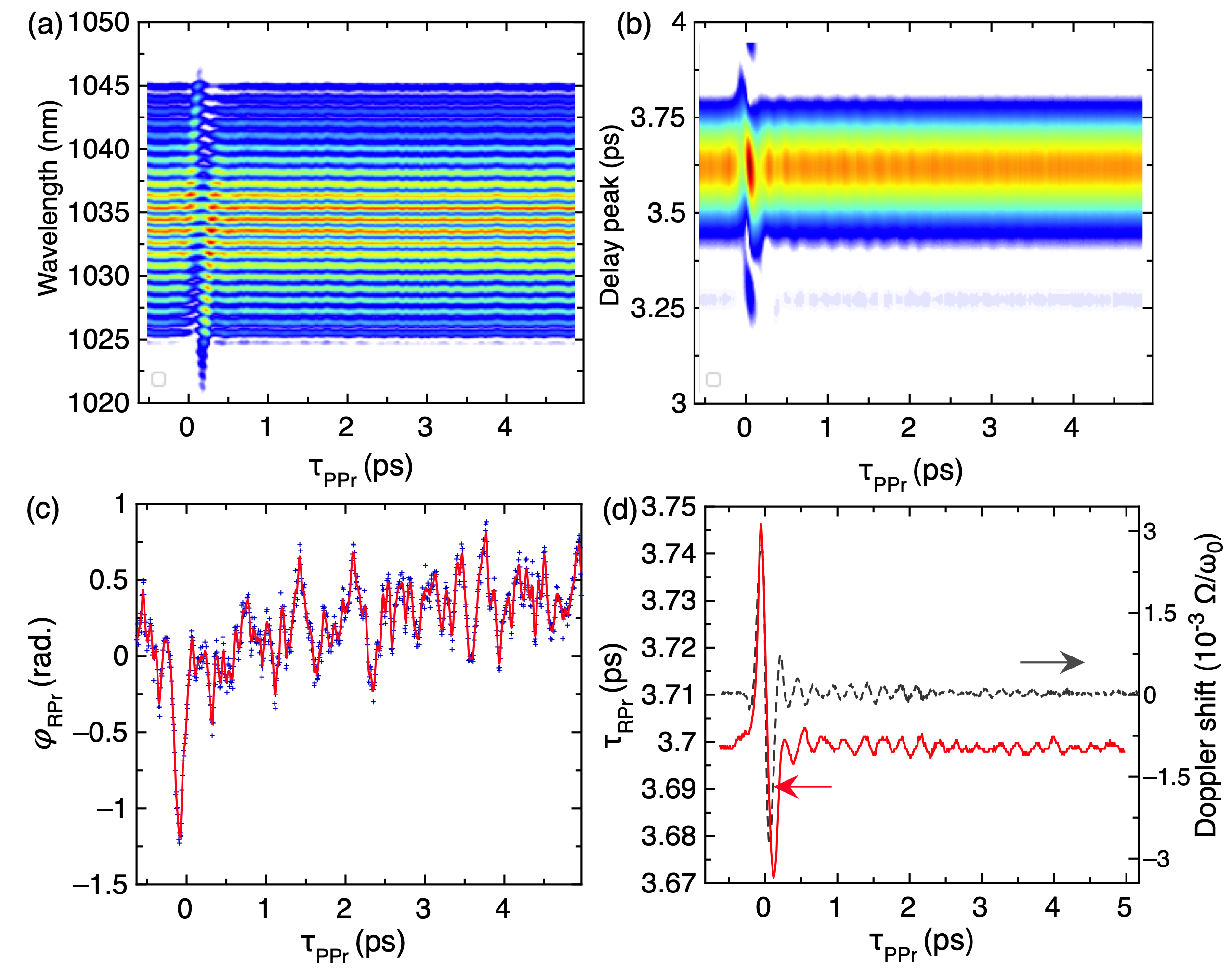}
\caption{Experimental data for the KTA crystal on a 6 ps scan range.
(a) Interferogram of the probe and reference pulses as a function of pump-probe delay.
(b) AC peak of the Fourier transform of the interferograms. 
(c) Relative phase between reference and probe pulses ($\varphi_\text{RPr}$) as a function of pump-probe delay. Raw data (blue), moving average (red).
(d) $\tau_\text{RPr}$ (red) and normalized Doppler shift (grey) as a function of pump-probe delay. 
}
\label{fig:kta}
\end{figure}

Although outside the range of validity of the model developed above, a proof-of-principle measurement show that delayed linear phenomena can also be investigated with our method. The temporal scanning range is then increased to +5\,ps, so as to evidence the delayed nonlinear answer of KTA (fast axis). The results are shown in Fig. \ref{fig:kta}. 

\begin{table}
\caption{\label{tab:table_kta}  Nonlinear refractive index at 1 $\mu$m of KTA ($\times10^{-20}\,\text{m}^{2}/\text{W}$), for 180 fs pulses.}
\begin{tabular}{lcc}
Sample & n$_{2,\text{slow}}$ & n$_{2,\text{fast}}$\\
\hline
KTA  & 7.8 $\pm$ 0.2  & 6.5 $\pm$ 0.2 \\
\end{tabular}
\end{table}

For this measurement, the pump energy is increased to 75\, $\mu$J to increase the signal-to-noise ratio at positive pump-probe delays. The group delay swing shows an asymmetry, followed by pseudo-periodic oscillations. This measurement evidences two main virbation modes, with respective wave numbers of 66\,cm$^{-1}$ and 75\,cm$^{-1}$, matching A1 vibrational frequencies of KTA, as reported in \cite{Tu:1996vr}. In case of a delayed nonlinear response, the signal seems mainly originating from temporal transcription of the phonon-induced spectral shift, as shown in Fig. \ref{fig:kta}d, where temporal and spectral oscillations perfectly match. This final measurement illustrates the potential application our method to low-frequency Raman spectroscopy \cite{Smith:2021wn}. 


\section{Conclusion}

To conclude, we have introduced a jitter-free time-resolved spectroscopic method to characterize the third-order nonlinearity on the femtosecond time-scale. 
This approach, coined as "nonlinear chirped Doppler interferometry", consists in monitoring the variation of the optical group delay of a transmitted probe under the effect of a strong pump pulse, rather than the phase, via spectral interferometry between the probe and a reference pulse sampled upstream. 
We have shown that the optical group delay transient changes originate from two different mechanism, coherently added up : (i) 2BC triggers energy exchanges between chirped pump and probe pulses and induces temporal reshaping; (ii) the XPM frequency-shift undergone by the probe is encoded in time by adequately chirping the reference pulse. 
Thanks to a good agreement between experimental data and a 1D numerical model, we have been able to define experimental working area of our method, in terms of chirp of involved pulses and nonlinear phase amount. 

Thus, we have demonstrated that monitoring this quantity instead of phase changes makes the detection insensitive to phase fluctuations and provides additional means to enhance the sensitivity and specificity of the detection, without resorting to heterodyne detection.
Nonlinear phase shifts as low as  10\,mrad, corresponding to a frequency shift of 30\,GHz, i.e. 0.01\% of the carrier frequency, can thus be detected. The method is suited to perform non-resonant $\chi^{(3)}$ spectroscopy in isotropic or anisotropic nonlinear media, but also  to survey resonant and delayed nonlinear processes. 
This original approach could be easily exported to the detection of spectral shifts driven by other linear or nonlinear processes, among them fluorescence, and Raman spectroscopy. 

Finally, the sensitivity reported here could be pushed further by multiple scans acquisition, higher averaging and extending the involved spectral bandwidths.

\section{Funding} We acknowledge financial support from the Agence Nationale de la Recherche France (Grant ANR19CE30000601  UNLOC), the European Regional Development Fund (OPTIMAL) and  from the European Union Horizon 2020 research and innovation program under the Marie Slodowska-Curie grant agreement No 860553.


\bibliographystyle{unsrt}

\bibliography{/Users/Aurele/Documents/Recherche/Bibliographie/bibtex}

\begin{thebibliography}{10}

\bibitem{Sutherland:2003aa}
R.~L. Sutherland.
\newblock {\em Handbook of nonlinear optics}.
\newblock Marcel Dekker INC, 2003.

\bibitem{Jullien200510-10-temporal-}
A.~Jullien, O.~Albert, F.~Burgy, G.~Hamoniaux, J.-P. Rousseau, J.-P. Chambaret,
  F.~Aug{\'e}-Rochereau, G.~Ch{\'e}riaux, J.~Etchepare, N.~Minkovski, and S.~M.
  Saltiel.
\newblock $10^{-10}$ temporal contrast for femtosecond ultraintense lasers by
  cross-polarized wave generation.
\newblock {\em Opt. Lett.}, 30(8):920--922, 2005.

\bibitem{Fattahi:2014aa}
H.~Fattahi, H.~G. Barros, M.~Gorjan, T.~Nubbemeyer, B.~Alsaif, C.~Y. Teisset,
  M.~Schultze, S.~Prinz, M.~Haefner, M.~Ueffing, A.~Alismail, L.~Vamos,
  A.~Schwarz, O.~Pronin, J.~Brons, X.~Tao Geng, G.~Arisholm, M.~Ciappina, V.~S.
  Yakovlev, D.-E. Kim, A.~M. Azzeer, N.~Karpowicz, D.~Sutter, Z.~Major,
  T.~Metzger, and F.~Krausz.
\newblock Third-generation femtosecond technology.
\newblock {\em Optica}, 1(1):45, 2014.

\bibitem{DeSalvo1996Infrared-to-ult}
R.~DeSalvo and E.~W.~Van Stryland.
\newblock Infrared to ultraviolet measurements of two-photon absorption and n2
  in wide bandgap solids.
\newblock {\em IEEE J. Quantum Electron.}, 32(8):1324, 1996.

\bibitem{Chien:00}
C.~Y. Chien, B.~La Fontaine, A.~Desparois, Z.~Jiang, T.~W. Johnston, J.~C.
  Kieffer, H.~P\'{e}pin, F.~Vidal, and H.~P. Mercure.
\newblock Single-shot chirped-pulse spectral interferometry used to measure the
  femtosecond ionization dynamics of air.
\newblock {\em Opt. Lett.}, 25(8):578--580, Apr 2000.

\bibitem{Chen:07}
Y.-H. Chen, S.~Varma, I.~Alexeev, and H.~M. Milchberg.
\newblock Measurement of transient nonlinear refractive index in gases using
  xenon supercontinuum single-shot spectral interferometry.
\newblock {\em Opt. Express}, 15(12):7458--7467, Jun 2007.

\bibitem{Borzsonyi:10}
\'{A}. B\"{o}rzs\"{o}nyi, Z.~Heiner, A.P. Kov\'{a}cs, M.~P. Kalashnikov, and
  K.~Osvay.
\newblock Measurement of pressure dependent nonlinear refractive index of inert
  gases.
\newblock {\em Opt. Express}, 18(25):25847--25854, Dec 2010.

\bibitem{Burgin2005Femtosecond-inv}
J.~Burgin, C.~Guillon, and P.~Langot.
\newblock Femtosecond investigation of the non-instantaneous third-order
  nonlinear suceptibility in liquids and glasses.
\newblock {\em Applied Physics Letters}, 87:211916, 2005.

\bibitem{LeBlanc:00}
S.~P.~Le Blanc, E.~W. Gaul, N.~H. Matlis, A.~Rundquist, and M.~C. Downer.
\newblock Single-shot measurement of temporal phase shifts by frequency-domain
  holography.
\newblock {\em Opt. Lett.}, 25(10):764--766, May 2000.

\bibitem{Thurston:2020aa}
R.~Thurston, M.~M. Brister, A.~Belkacem, T.~Weber, N.~Shivaram, and D.~S.
  Slaughter.
\newblock Time-resolved ultrafast transient polarization spectroscopy to
  investigate nonlinear processes and dynamics in electronically excited
  molecules on the femtosecond time scale.
\newblock {\em Review of Scientific Instruments}, 91:053101, 2020.

\bibitem{Kang:97}
Inuk Kang, Todd Krauss, and Frank Wise.
\newblock Sensitive measurement of nonlinear refraction and two-photon
  absorption by spectrally resolved two-beam coupling.
\newblock {\em Opt. Lett.}, 22(14):1077--1079, Jul 1997.

\bibitem{Smolorz:00}
Sylvia Smolorz and Frank Wise.
\newblock Femtosecond two-beam coupling energy transfer from raman and
  electronic nonlinearities.
\newblock {\em J. Opt. Soc. Am. B}, 17(9):1636--1644, Sep 2000.

\bibitem{Smolorz:1999vf}
S.~Smolorz, F.~Wise, and N.~F. Borelli.
\newblock Measurement of the nonlinear optical response of optical fiber
  materials by use of spectrally resolved two-beam coupling.
\newblock {\em Opt. Lett.}, 24(16):1103, 1999.

\bibitem{Wahlstrand:2013wl}
J.~K. Wahlstrand, J.~H. Odhner, E.~T. McCole, Y.-H. Cheng, J.~P. Palestro,
  R.~J. Levis, and H.~M. Milchberg.
\newblock Effect of two-beam coupling in strong-field optical pump-probe
  experiments.
\newblock {\em Phys. Rev. A}, 87:053801, 2013.

\bibitem{Patwardhan:21}
Gauri~N. Patwardhan, Jared~S. Ginsberg, Cecilia~Y. Chen, M.~Mehdi Jadidi, and
  Alexander~L. Gaeta.
\newblock Nonlinear refractive index of solids in mid-infrared.
\newblock {\em Opt. Lett.}, 46(8):1824--1827, Apr 2021.

\bibitem{Boyd1992Nonlinear-optic}
R.~W. Boyd.
\newblock {\em Nonlinear optics}.
\newblock Academic Press, INC, 1992.

\bibitem{Ripoche:1997aa}
J.~F. Ripoche, G.~Grillon, B.~Prade, M.~Franco, E.~Nibbering, R.~Lange, and
  A.~Mysyrowicz.
\newblock Determination of the time dependence of n2 in air.
\newblock {\em Optics Communications}, 135:310, 1997.

\bibitem{Ghosh:99}
G.~Ghosh.
\newblock Dispersion-equation coefficients for the refractive index and
  birefringence of calcite and quartz crystals.
\newblock {\em Opt. Commun.}, 163:95--102, 1999.

\bibitem{Malitson:72}
I.~H. Malitson and M.~J. Dodge.
\newblock Refractive index and birefringence of synthetic sapphire.
\newblock {\em J. Opt. Soc. Am.}, 62:1405, 1972.

\bibitem{Kabacinski:19}
Piotr Kabaci\'{n}ski, Tomasz~M. Karda\'{s}, Yuriy Stepanenko, and Czes{\l}aw
  Radzewicz.
\newblock Nonlinear refractive index measurement by spm-induced phase
  regression.
\newblock {\em Opt. Express}, 27(8):11018--11028, Apr 2019.

\bibitem{Adair1989nonlinear-refra}
R.~Adair, L.~L. Chase, and S.~A. Payne.
\newblock Nonlinear refractive index of optical crystals.
\newblock {\em Physical Review B}, 39(5):3337--3350, 1989.

\bibitem{Major:04}
A.~Major, F.~Yoshino, I.~Nikolakakos, J.~S. Aitchison, B.~Lavorel, and P.~W.~E.
  Smith.
\newblock Dispersion of the nonlinear refractive index in sapphire.
\newblock {\em Opt. Lett.}, 29(29):602--604, May 2004.

\bibitem{Minkovski2004Nonlinear-polar}
N.~Minkovski, G.~I. Petrov, S.~M. Saltiel, O.~Albert, and J.~Etchepare.
\newblock Nonlinear polarization rotation and orthogonal polarization
  generation experienced in a single-beam configuration.
\newblock {\em J. Opt. Soc. Am. B}, 21(9):1659--1664, 2004.

\bibitem{Canova2008Efficient-gener}
L.~Canova, S.~Kourtev, N.~Minkovski, A.~Jullien, R.~Lopez-Martens, O.~Albert,
  and S.~M. Saltiel.
\newblock Efficient generation of cross-polarized femtosecond pulses in cubic
  crystals with holographic cut orientation.
\newblock {\em Appl. Phys. Lett.}, 92:231102, 2008.

\bibitem{Kourtev:09}
S.~Kourtev, N.~Minkovski, L.~Canova, A.~Jullien, O.~Albert, and S.~M. Saltiel.
\newblock Improved nonlinear cross-polarized wave generation in cubic crystals
  by optimization of the crystal orientation.
\newblock {\em J. Opt. Soc. Am. B}, 26(7):1269--1275, 2009.

\bibitem{Gubler:00}
U.~Gobler and C.~Bosshard.
\newblock Optical third-harmonic generation of fused silica in gas atmosphere:
  Absolute value of the third-order nonlinear optical susceptibility.
\newblock {\em Phys. Rev. B}, 61(7):10702, 2000.

\bibitem{Li:2001tm}
H.~P. Li, C.~H. Kam, Y.~L. Lam, and W.~Ji.
\newblock Femtosecond z-scan measurements of nonlinear refraction in nonlinear
  optical crystals.
\newblock {\em Optical Materials}, 15:237--242, 2001.

\bibitem{Tu:1996vr}
C.~Tu, A.~R. Guo, R.~Tao, R.~S. Katiyar, R.~Guo, and A.~S. Bhalla.
\newblock Temperature dependent raman scattering in ktiopo4 and ktioaso4 single
  crystals.
\newblock {\em Journal of Applied Physics}, 79:3235, 1996.

\bibitem{Smith:2021wn}
D.~R. Smith, J.~J. Field, D.~G. Winters, S.~R. Domingue, F.~Rininsland, D.~J.
  Kane, J.~W. Wilson, and R.~A. Bartels.
\newblock Phase noise limited frequency shift impulsive raman spectroscopy.
\newblock {\em APL Photonics}, 6:026107, 2021.

\end{thebibliography}

\end{document}